\documentclass[letter,desactivate]{aa}  

\DeclareRobustCommand{\VAN}[3]{#2}
\let\VANthebibliography\thebibliography
\def\thebibliography{\DeclareRobustCommand{\VAN}[3]{##3}\VANthebibliography}

\usepackage[colorlinks=true,
    linkcolor=blue, citecolor=blue, filecolor=blue, urlcolor=blue]{hyperref}
\usepackage{graphicx}	
\usepackage{amsmath}	
\usepackage{amssymb}	
\usepackage{array}
\usepackage{multirow}
\usepackage{bigdelim}
\usepackage[varg]{txfonts}
\usepackage{xcolor}
\usepackage[version=4]{mhchem}

\newcommand{\NHnH}{$N_{\rm eff}(\rm H) - n_{\rm H}\,$}

\defcitealias{Bisbas2023}{B23}

\begin{document}

\title{An analytic formalism to describe the \NHnH relationship in molecular clouds}
\titlerunning{}
\authorrunning{B.A.L. Gaches}

\author{Brandt A. L. Gaches
        \inst{1}\thanks{E-mail: brandt.gaches@uni-due.de}
        }
\institute{
Faculty of Physics, University of Duisburg-Essen, Lotharstraße 1, 47057 Duisburg, Germany
}

\date{Accepted XXX. Received YYY; in original form ZZZ}

\abstract
{Astrochemical modeling requires, as input, the effective column density of gas (or extinction) that attenuates an external, isotropic, far-ultraviolet radiation field. In three-dimensional simulations, this can be calculated through ray-tracing schemes, while in 0D chemical models it is often treated as a free parameter.}{We aim to produce an analytic, physically motivated formalism to predict the average relationship between the effective hydrogen-nuclei column density, $N_{\rm eff}({\rm H})$, and the local hydrogen-nuclei number density, $n_{\rm H}$.}{We construct an analytic model utilizing characteristic length scales that connects the turbulence-dominated regime and the gravitational-dominated regime at high-density.}{The model well-reproduces a previous analytic fit to simulation results and is consistent with the high-density power-law indices, e.g., $N_{\rm eff}(H) \propto n^{\gamma}$, of $\gamma \approx 0.4 - 0.5$ found in previous numerical simulations utilizing ray-tracing.}{We present an analytic model relating the average effective column density, $N_{\rm eff}$, to the local number density, $n_{\rm H}$, which reproduces the behaviors found in three-dimensional simulations. The analytic model can be utilized as a sub-grid prescription for shielded molecular gas or in astrochemical models for a physically motivated estimation of the attenuating column density.}


\maketitle

\section{Introduction}
The chemistry of the molecular clouds plays a vital role in our understanding of the star formation process. Observationally, the dynamics of molecular clouds are constrained through the Doppler shift of molecular line emission. Chemistry also plays an important role in the thermodynamics of molecular clouds through atomic and molecular line cooling \citep{Draine2011, Tielens2013}. Therefore, astrochemical models of molecular clouds are central in both theoretical and observational studies of star formation and the physics of molecular clouds. 

In astrochemical modeling, the effective attenuating column density for the external radiation field is a primary input \citep{Wakelam2013, Bovino2024}. This effective column density, $N_{\rm eff}$ (column density of hydrogen nuclei), or effective extinction, $A_{\rm V, eff}$ (magnitude), gives the local FUV flux, $\chi/\chi_0 \propto e^{-A_{\rm V, eff}}$, where $\chi_0$ is the FUV flux at the cloud boundary. In the case of isotropic radiation, the effective column density at a point, 
$\bf{x}$, is the weighted average (see below) of all directions of the integrated column density from the point $\bf{x}$ to the cloud boundary. It is thus different from the observational column density, which is integrated through the entire cloud along an observed line of sight. Given the importance of this parameter to astrochemical models, estimating it is crucial for accurate chemical models of molecular clouds.

Astrochemical models have to make a choice for $N_{\rm eff}$, such as assuming it is equal to the observed column density, or maintaining it as a free parameter in model fitting. In molecular cloud simulations, $N_{\rm eff}$ is often calculated through ray-tracing \citep[e.g.,][]{Nelson1997, Glover2007, Glover2010, VanLoo2013, Safranek-Shrader2017, Seifried2017, Hu2021, Wunsch2024}, where the hydrogen nuclei column density is summed along each ray out of the domain from each cell using, e.g., a {\sc healpix}-based method for the ray directions, or the simpler six-ray approach along each cardinal direction. The effective extinction is the average, 
\begin{equation}
    A_{\rm V,eff} = -\frac{1}{\gamma} \log \left (\frac{1}{N_{\rm rays}} \sum_i^{N_{\rm rays}} e^{-\gamma A_{V,i}} \right ),
\end{equation}
where the coefficient, $\gamma$, is a normalization constant chosen to be the photo-destruction rate coefficient for particular molecules \citep[e.g.,][]{Glover2010, Hu2021} or based on dust absorption \citep[e.g.,][]{Nelson1997}. When no ray-tracing is used, $N_{\rm eff}$ is computed using an assumed physical length scale and the cell's density (see discussion in \citet{Safranek-Shrader2017}). Recently, \citet{Bisbas2023} (hereafter \citetalias{Bisbas2023}) provided a fit to a several hydrodynamical simulations for $A_{\rm V, eff}-n_{\rm H}$,
\begin{equation}\label{eq:B23}
    A_{\rm V, eff} = 0.05 \exp\left [1.6\, \left ( \frac{n_{\rm H}}{\rm cm^{-3}}  \right )^{0.12} \right ].
\end{equation}
However, this fit is biased by the assumptions and initial conditions in the underlying simulations: the simulations were predominantly for Milky Way-like clouds, with both a resolution limit and a numerical column density floor. \citet{Safranek-Shrader2017} and \citet{Hu2021} also fitted a polynomial from their simulations and for $n_{\rm H} > 10$~cm$^{-3}$ found $A_V \propto n_{\rm H}^{0.40}$.

In this Letter, we present a simple analytic model connecting the local gas density to the effective column density. The model is derived such that, with few assumptions about the cloud-scale turbulence and dynamics, an effective column density can be estimated that is in general agreement with the results from three-dimensional simulations. In Section \ref{sec:method} we present the underlying model components. In Section \ref{sec:res}, we show the results of the model in comparison to the previous three-dimensional calculations. Finally, in \ref{sec:discconc}, we discuss the potential applications for this model, its limitations, and concluding remarks.

\section{Methods}\label{sec:method}
\begin{figure}
    \centering
    \includegraphics[trim={1.5cm 1.25cm 1cm 0.75cm}, clip, width=0.49\textwidth]{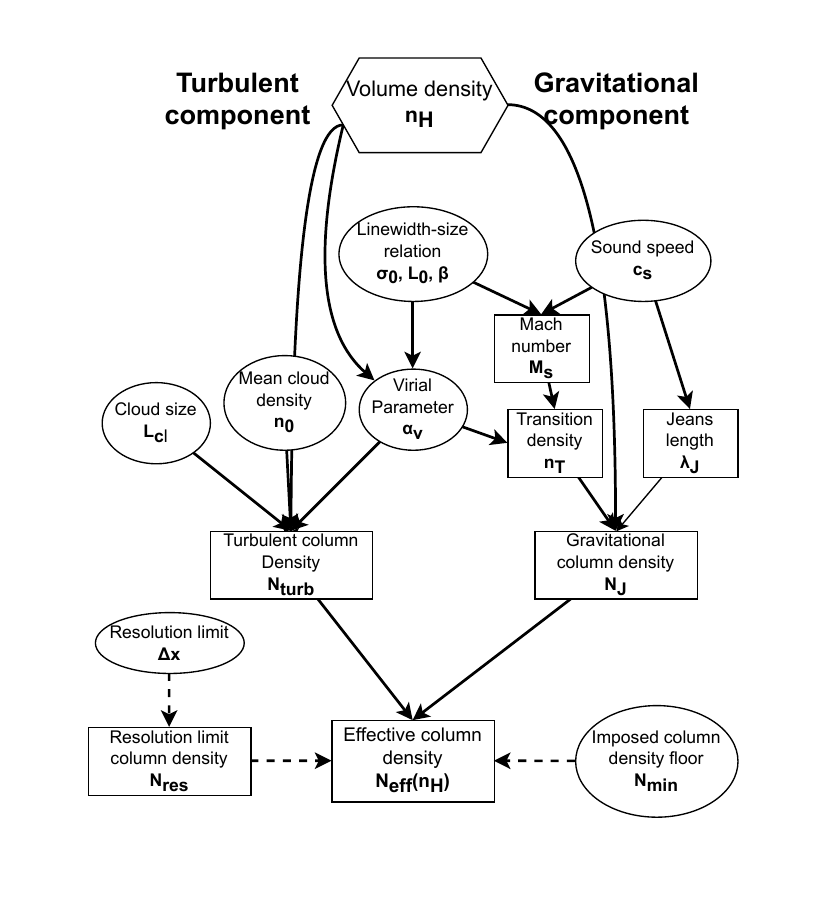}
    \caption{\label{fig:schematic}Chart showing the different components and where parameters enter into the model.}
\end{figure}

We build an analytic model to reproduce a characteristic \NHnH relationship from a simple, physically intuitive construction. The model is built up by considering three different characteristic length scales: i) turbulent, ii)  gravitational, and iii) numerical resolution. Figure \ref{fig:schematic} shows a chart of the scales of interest and the various parameters that enter into the model.

\subsection{Turbulence length scale}\label{sec:turbcol}
Within the molecular cloud, any particular parcel of gas is surrounded by an envelope of turbulent gas that attenuates the external FUV radiation. We assumed this column density is, on average, related to an effective length scale, $\ell_{\rm turb}$, out of the cloud through the turbulent envelope multiplied by the average number density, $n_0$, in the turbulent envelope,
\begin{equation}\label{eq:Nturb}
    N_{\rm turb} \approx \ell_{\rm turb} n_0.
\end{equation}
If the cloud depth is known, e.g., from three-dimensional cloud maps using observations from Gaia \citep{Zucker2021, RezaeiKh2024, Edenhofer2024}, and there is a measured column density distribution function (N-PDF), this can be estimated by dividing the mean of the log-normal component by the cloud depth. 

We assumed that at number density, $n_{\rm H}$, we can determine a characteristic length scale $\ell$, given assumptions on the macroscopic cloud properties. We utilized the virial parameter, $\alpha_V$, to calculate this effective size as a function of density, defined by
\begin{equation}\label{eq:virial}
    \alpha_V = \frac{5\sigma(\ell)^2}{G\mu_H m_H n_{\rm H} \ell^2},
\end{equation}
where $\sigma(\ell)$ is the turbulence velocity dispersion as a function of the length scale, $G$ is the gravitational constant, $\mu_H$ is the mean molecular weight per hydrogen nucleus, and $m_H$ is the mass of hydrogen. To get the above expression, we have assumed that the mass and density are connected via $M \approx \rho \ell^3$. Relating a length scale to the density and virial parameter assumes a hierarchical density structure. The linewidth-size relationship is
\begin{equation}\label{eq:linesize}
    \sigma(\ell) = \sigma_0 \left ( \frac{\ell}{L_0}\right )^\beta,
\end{equation}
where $\sigma_0$ and $L_0$ are the normalization velocity and scale (or driving scale), respectively, and $\beta$ is related to the physics behind the turbulence cascade \citep{Larson1981, Heyer2015}  Recent surveys have shown that $\beta$ typically ranges between 0.3 - 0.67 \citep{Rice2016, Kauffmann2017, Spilker2022}. Combining Eqs. \ref{eq:virial} and \ref{eq:linesize} leads to the length scale, 
\begin{equation}
    \ell = \left [ \frac{G L_0^{2\beta}\mu_H m_H n_{\rm H}}{5\sigma_0^2} \alpha_v\right ]^{1/[2(\beta-1)]}.
\end{equation}

At low densities or high virial parameters, this length scale, $\ell$, may exceed the cloud size, $L_{\rm cl}$. Therefore, we computed a modified length to limit $\ell$ to be less than $L_{\rm cl}$, 
\begin{equation}\label{eq:leff}
    \bar{\ell} = \ell \times S_\ell(z_{\ell}) 
\end{equation}
where $z_\ell = L_{\rm cl}/\ell$. The dampening function requires that for $\ell << L_{\rm cl}$,  $\bar{\ell} \rightarrow \ell$,  and $\ell \gtrsim L_{\rm cl}$, $\bar{\ell} \rightarrow L_{\rm cl}$. We adopted $S_\ell(z_\ell) = {\rm erf}(\frac{\sqrt{\pi}}{2}z_\ell)$, since it well approximates a step-function filter, while remaining smooth at all scales. This choice primarily impacts the lowest densities. The length scale out of the cloud is then
\begin{equation}
\ell_{\rm turb} = R_{\rm cl} - \frac{1}{2}\bar{\ell},
\end{equation}
where $R_{\rm cl} = \frac{1}{2}L_{\rm cl}$ is the cloud radius. In low-density gas, more likely to be in the outer regions, $\ell_{\rm turb} \rightarrow 0$, and so $N_{\rm turb} \rightarrow 0$. Conversely, in the dense gas, $\ell_{\rm turb} \rightarrow R_{\rm cl}$, and so
$N_{\rm turb} \rightarrow  n_0 R_{\rm cl}$. Since the effective column density for an isotropic external radiation field is most sensitive to the lowest column density out of the cloud, it can be readily seen for any approximately spherically symmetric cloud that the maximum $\ell_{\rm turb}$ is $R_{\rm cl}$. 

\subsection{Gravitational length scale}\label{sec:gravcol}
In the high-density limit, the effective column density is dominated by gravitationally bound gas in the immediate surroundings. The length scale of relevance here is the Jeans length, $\lambda_J$. We defined this component as 
\begin{equation}\label{eq:jeans}
    N_J(n_{\rm H}) \approx \frac{1}{2}\lambda_J n_{\rm H} = \frac{1}{2}\left ( \frac{\pi}{G\mu_H m_H} \right )^{1/2} c_s n_{\rm H}^{1/2},
\end{equation}
where $c_s = \sqrt{k_B T/\mu_p m_H}$ is the sound speed, $T$ is the gas temperature, $k_B$ is the Boltzmann constant, and $\mu_p$ is the mean molecular mass per free particle. Under isothermal conditions, $N_{J} \propto n_{\rm H}^{1/2}$. 

We accounted for the fact that gravity is not dominant at all densities by introducing a threshold that determines when gravity becomes dynamically important. We denoted the threshold function as $S_n(n_{\rm H}/n_T)$, where $n_T$ is the transition density. A good parameterization of $n_T$ equates the Jeans length and turbulence sonic scale, and following \citet{Burkhart2019} for isothermal turbulence,
\begin{equation}\label{eq:nt}
    \frac{n_T}{n_0} = \frac{\pi^2}{15}\alpha_V M_s^2,
\end{equation}
where $M_s$ and $\alpha_V$ calculated at $L_0$ using $\sigma_0$, which is a similar expression also shown by \citet{Padoan2011}.

We adopted a softened threshold, $S_n(z_n) = z_n/(1+z_n)$, where $z_n = n_{\rm H}/n_T$. The choice of threshold will impact how rapidly or slowly the effective column density transitions between the turbulence and gravity-dominated regimes. The total gravitational column density is 
\begin{equation}
    N_{J,T}(n_{\rm H}) =  S_n(n_{\rm H}/n_T) \times N_J.
\end{equation}

\subsection{Resolution limits}\label{sec:reslim}
In hydrodynamic simulations, the total attenuating column is constrained by the limited resolution \citep{Glover2010, VanLoo2013}. The resolution limit imposes a floor for the attenuation column density,
\begin{equation}
    N_{\Delta x} \approx C_{\Delta x} n_{\rm H} \Delta x,
\end{equation}
where the prefactor $C_{\Delta x}$ depends on how the self-contribution is accounted for, e.g., a whole cell size or half a cell size.

\subsection{Combined model}\label{sec:totalmod}
The final \NHnH relation without numerical resolution effects is
\begin{equation}\label{eq:NHnH}
    N_{\rm eff}(n_{\rm H}) = N_{\rm turb}(n_{\rm H}) + N_{J,T}(n_{\rm H}).
\end{equation}
If there is a needed restriction based on resolution, then the expression is modified
\begin{equation}\label{eq:NHnH_res}
    N_{\rm eff}(n_{\rm H}) = \max(N_{\rm turb}(n_{\rm H}) + N_{J,T}(n_{\rm H}), N_{\Delta x}(n_{\rm H})).
\end{equation}
For comparison to \citetalias{Bisbas2023}, we also used a floor of 0.05 mag.

\section{Results}\label{sec:res}
For our fiducial model, we aim to model Milky Way-like molecular clouds, and adopt the fiducial parameters in Table \ref{tab:params}, mean molecular weights, $\mu_H = 1.4$ and $\mu_p = 2.33$, a conversion factor between the hydrogen nuclei column density and visual extinction, $A_V = 6.29\times10^{-22} N(\ce{H})$ \citep{Rollig2007}, and a normalization scale for the linewidth-size relationship of $L_0 = 1$ pc. In Appendix \ref{app:cs_mu}, we investigate the impact of assuming a fixed $c_s$ and $\mu_p$ and show that assumptions on $\mu_p$ have a minimal impact while a variable $c_s$ can increase $N_{\rm eff}$ at intermediate and low densities. However, these have less of an impact compared to varying key physical parameters.
\begin{table}
    \caption{Important physical parameters}              
    \label{tab:params}     
    \centering
    \begin{tabular}{c|c|c}          
    \hline
    Parameter & Fiducial & Range \\    
    \hline 
        $L$ (pc) & 15 & 1 - 20\\
        $T$ (K) & 10 & 10 - 100\\
        $n_0$ (cm$^{-3}$) & 100 & 10 - 1000\\
        $\alpha_V$ & 1 & 0.5 - 5\\
        $\sigma_0$ (km s$^{-1}$) & 0.38$^{1}$ & 0.2 - 5.0 \\
        $\beta$ & 0.5$^{1}$ & 0.3 - 0.7 \\
    \hline 
    \end{tabular}
    \tablebib{(1)~\citet{Rice2016}}
\end{table}

\begin{figure}
    \centering
    \includegraphics[width=\columnwidth]{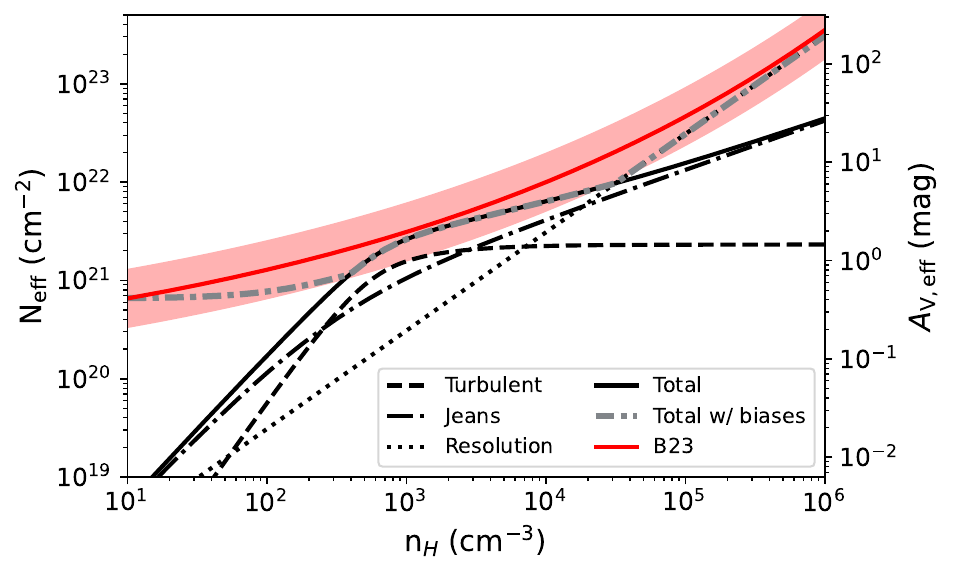}
    \caption{Fiducial model results for the \NHnH relationship. The different components and totals are annotated. The red line shows the \citet{Bisbas2023}fit, and the band shows a factor of 2 deviation away. The secondary y-axis shows the extinction using a scalar conversion factor as described in the text. ``Total w/ biases'' includes the resolution limit and floor.}
    \label{fig:NHnH_fid}
\end{figure}

Figure \ref{fig:NHnH_fid} shows the fiducial model in comparison to the \citetalias{Bisbas2023} fit. At moderate densities, the model matches \citetalias{Bisbas2023} within a factor of two, although towards low density, a floor is needed to remain consistent with \citetalias{Bisbas2023}. The slope flattens around $n_{\rm H} \approx 500$ cm$^{-3}$ due to the transition between the turbulent to gravitational components. At high density, $N_{\rm eff}$ is dominated by the gravitational component with a slope flatter than 0.5 due to the contribution of the turbulent column, even with an isothermal assumption. The slope of the fiducial model between $10^3 \leq n_{\rm H} \leq 10^5$ is consistent with the $\propto n_{\rm H}^{0.42}$ and $\propto n_{\rm H}^{0.4}$ found in the simulations of \citet{Safranek-Shrader2017} and \citet{Hu2021}, respectively. The inclusion of the relevant resolution limit and floor brings the model into excellent agreement with \citetalias{Bisbas2023}.

\begin{figure*}[tb!]
    \centering 
    \includegraphics[width=\textwidth]{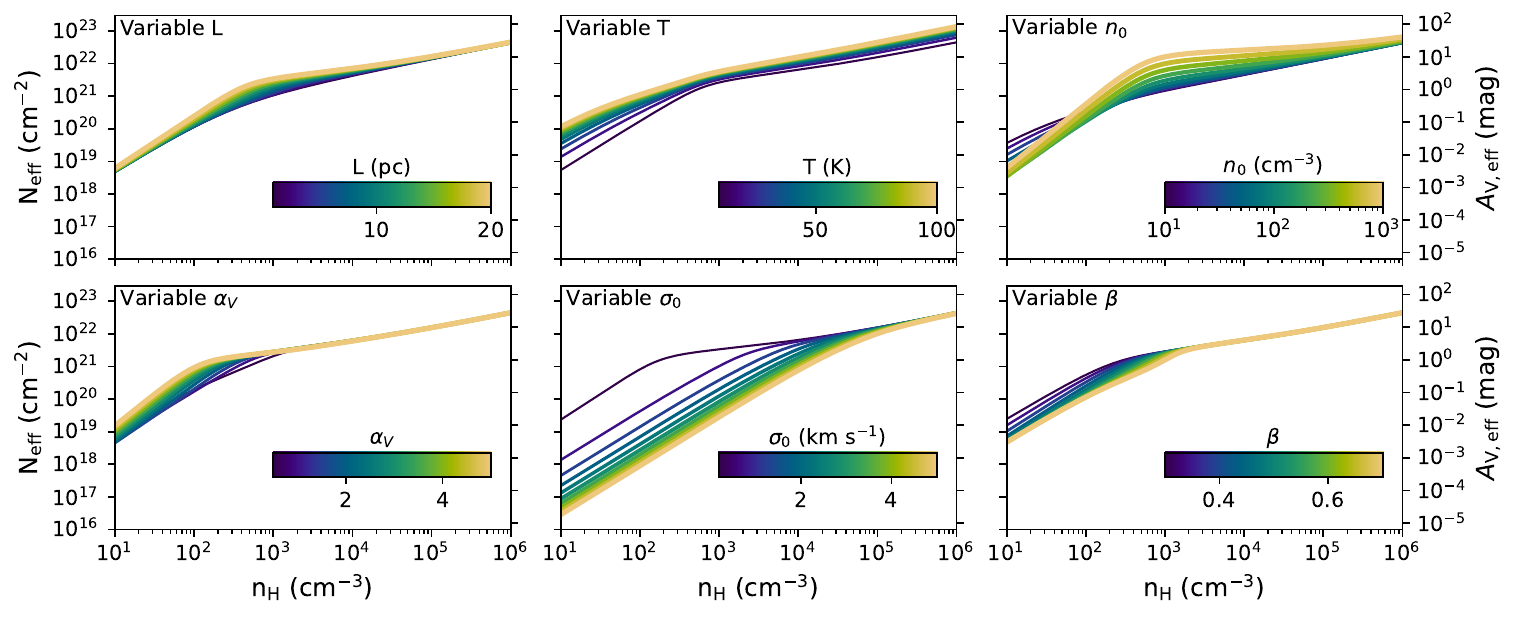}
    \caption{Parameter space results for the \NHnH relationship. Each panel varies the annotated variable while keeping the rest fixed to the fiducial values shown in Table \ref{tab:params}.}
    \label{fig:NHnH_param}
\end{figure*}

We also investigated the impact of varying the major underlying parameters in Table \ref{tab:params}. Figure \ref{fig:NHnH_param} shows the impact of varying six different parameters on the \NHnH relationship. For the parameter study, we only varied the described parameter while keeping the rest fixed at the fiducial value. At high density, the only parameters that have a noticeable impact are the gas temperature, which increases the Jeans radius, and the average ambient turbulent density, which globally shifts the attenuating column. 

As expected, increasing the cloud size or the average turbulent density increases the turbulent column and causes the slope to flatten around $n({\rm H}) \approx 10^3$ cm$^{-3}$. Further, decreasing the strength of the turbulence or the linewidth-size slope increases the turbulent column density. For isotropic hydrodynamic turbulence, the \NHnH relationship maintains a relatively constant slope at all densities, consistent with the $N_{\rm eff} \propto n^{0.42}$ found in the hydrodynamic simulations of \citet{Safranek-Shrader2017}. The virial parameter does not have a significant impact on the results, since it impacts both $\ell$ and $n_T$ in opposite directions. In Appendix \ref{app:plaw}, we detail the calculation of the slopes of the \NHnH function and comparison to the fitted power-law slopes and show that towards high densities the model is in good agreement with simulations.

\section{Discussion and Conclusions}\label{sec:discconc}
An analytic expression for the \NHnH relationship is advantageous for both simulations and chemical models of molecular clouds. Astrochemical codes, in particular 0D models, rely on the input of a meaningful attenuating column density along with the density, temperature, and surface FUV radiation field. Further, the \NHnH relationship can be used to generate a one-dimensional density distribution for rapid one-dimensional photodissociation region models \citep[see, e.g.,][]{Bisbas2021, Bisbas2023}. For observations of unresolved molecular clouds, the use of \NHnH one-dimensional photodissociation region models can be used with the input physical parameters to get insights into the physics of the cloud.

In hydrodynamic simulations modelling molecular clouds, performing radiative transfer may be cost-prohibitive. In such instances, this analytic model can be utilized as a subgrid model to rapidly estimate the attenuating column density. While this will not replicate the accuracy of a full radiative transfer simulation, it is worth highlighting that even in such simulations, the spread in $A_v$ as a function of density decreases with increasing density (see e.g. \citet{Hu2021}), so towards higher density, the impact of using this model versus full raytracing is reduced. Finally, attenuating column densities are not just used for FUV radiation, but also for cosmic-ray energy losses into clouds \citep[see, e.g.,][]{Padovani2009, Gaches2022}.

In this Letter, we have presented an analytic model to estimate the effective column density, $N_{\rm eff}$, which acts as the attenuating column density of an isotropic external radiation field, as a function of the local hydrogen-nuclei number density, $n_{\rm H}$. We demonstrated that our fiducial model, which aims to represent clouds similar to the solar neighborhood, qualitatively reproduces the function presented by \citetalias{Bisbas2023}, which is a fit to several three-dimensional simulations with radiative transfer, and the power-law behavior towards high densities found in other works, such as \citet{Glover2010}, \citet{VanLoo2013}, \citet{Safranek-Shrader2017}, and \citet{Hu2021}. The analytic model can enable rapid estimation of effective column densities for astrochemical models and act as a subgrid model in simulations of molecular clouds that do not include radiative transfer. Future work will combine this model into grids of photodissociation region codes to estimate macroscopic cloud parameters from molecular line observations.

\begin{acknowledgements}
We thank the anonymous referee for their useful suggestions, which improved this Letter, and Thomas Bisbas and Daniel Seifried for their thoughtful discussions. BALG is supported by the German Research Foundation (DFG) in the form of an Emmy Noether Research Group - DFG project \#542802847 (GA 3170/3-1).
\end{acknowledgements}

\bibliographystyle{aa}
\bibliography{lib} 

\onecolumn
\begin{appendix}
\section{Impact of density-dependent temperature and $\mu_p$}\label{app:cs_mu}
Our model presented in the main text only considered constant temperatures and mean molecular mass per free particle, resulting in constant sound speeds. This is clearly an oversimplification, so we investigated the impact of a density-dependent sound speed, $c_s(n_{\rm H})$, and mean molecular mass per free particle, $\mu_p(n_{\rm H})$, on the calculated \NHnH relationship. We used the 1D $A_V - n_{\rm H}$ density distribution from \citetalias{Bisbas2023} (Eq. \ref{eq:B23}) with the public photodissociation code {\sc 3d-pdr} \citep{Bisbas2012}, including a total H$_2$ cosmic-ray ionization, $\zeta_2 = 10^{-16}$ s$^{-1}$ and an external FUV field of $\chi = 1$ in the unit of the \citet{Draine1978} field. We used a reduced chemical network of 33 species and 331 reactions, with initial abundances to reproduce Milky Way-like molecular clouds \citep[see][for details]{Bisbas2012}. 

\begin{figure*}[htb!]
    \centering
    \includegraphics[width=\textwidth]{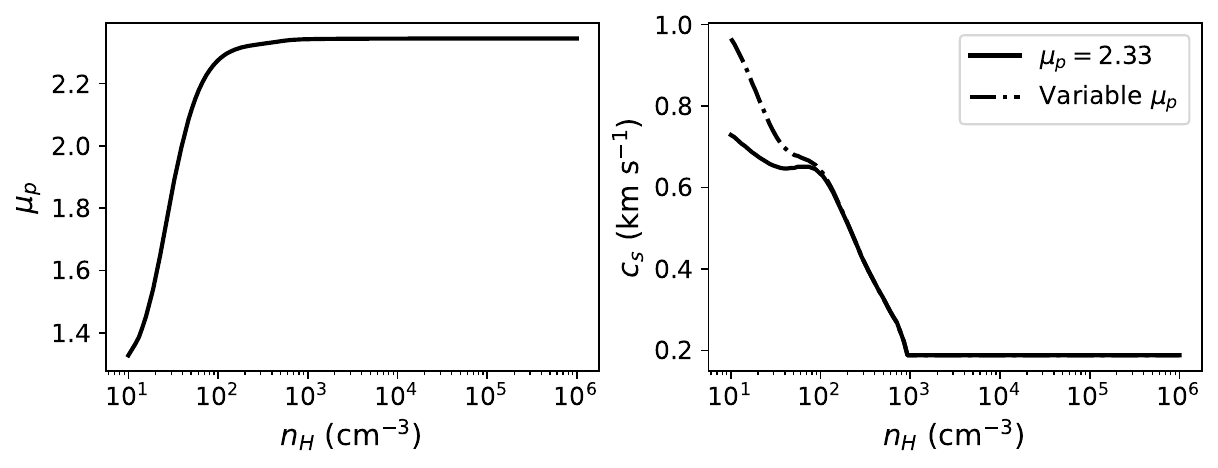}
    \caption{\label{fig:csmu}Variation of $\mu_p$ (left) and $c_s$ (right) with number density, $n_{\rm H}$, using one-dimensional models from {\sc 3d-pdr}. The figure shows $c_s$ assuming both a constant $\mu_p$ and a variable $\mu_p$.}
\end{figure*}

This is not a self-consistent treatment, but since the \NHnH function with constant $c_s$ and $\mu_p$ is similar to the \citetalias{Bisbas2023} fit, the use of the PDR model provides a qualitative understanding of the impact of variable $c_s$ and $\mu_p$. We compute $c_s(n_{\rm H})$ and $\mu_p(n_{\rm H})$ from the gas temperature and chemical abundances directly. Figure \ref{fig:csmu} shows the computed $\mu_p(n_{\rm H})$ and $c_s(n_{\rm H})$, where we have shown $c_s$ calculated with a constant and a variable $\mu_p$.

\begin{figure*}[htb!]
    \centering
    \begin{tabular}{ccc}
        \includegraphics[width=0.33\textwidth]{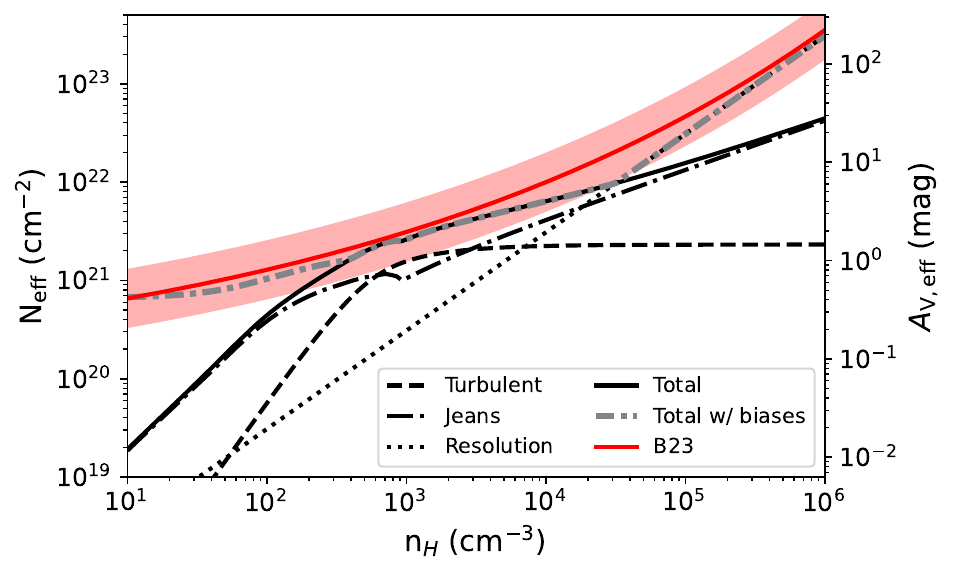} & 
        \includegraphics[width=0.33\textwidth]{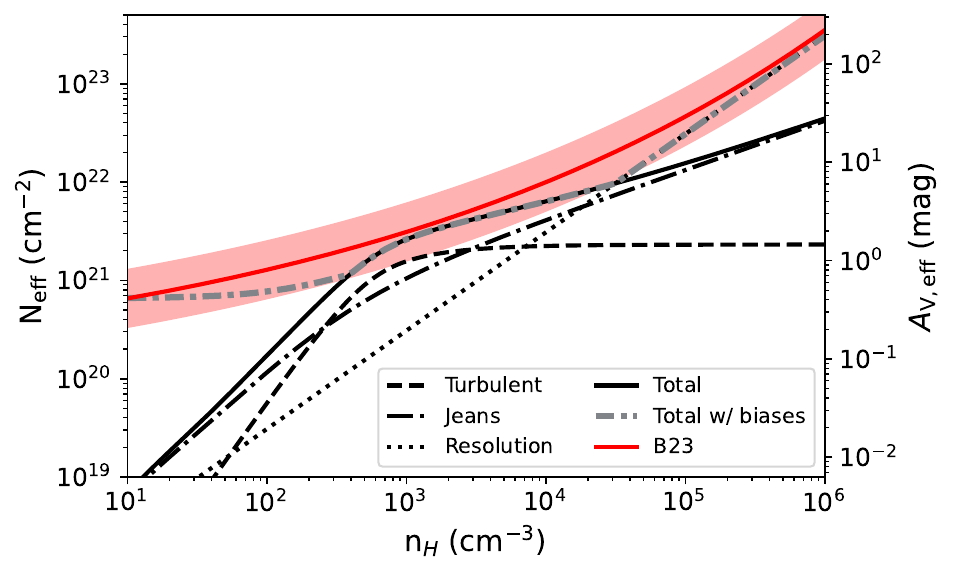} &
        \includegraphics[width=0.33\textwidth]{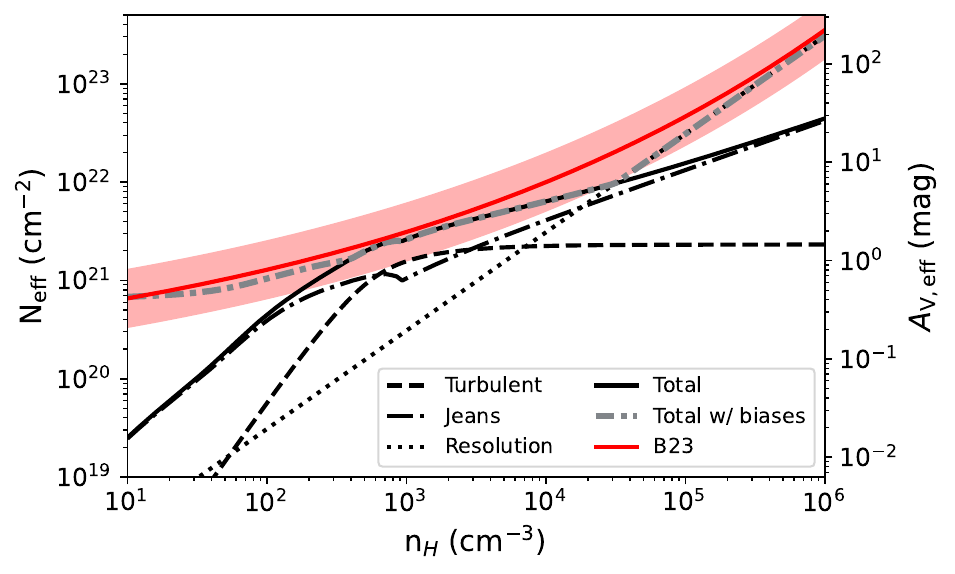} \\
        a)$c_s(n_{\rm H})$ & b)$\mu_p(n_{\rm H})$ & c)$c_s(n_{\rm H}), \mu_p(n_{\rm H})$
    \end{tabular}
    \caption{\label{fig:NHnH_cs_mu}\NHnH relationship modifed by using a variable $c_s(n_{\rm H})$ (left), $\mu_p(n_{\rm H})$ (middle), and both $c_s(n_{\rm H})$ and $\mu_p(n_{\rm H})$ (right), as annotated below each subplot.}
\end{figure*}

Figure \ref{fig:NHnH_cs_mu} shows the impact of a density-dependent $c_s$ and $\mu_p$. Varying $\mu_p$ has a marginal impact: it enters weakly into the Jeans length and only deviates from a constant $2.33$ in regimes where the gravitational component is anyway negligible. The impact of a variable sound speed is most noticeable at lower densities, increasing the gravitational component due to two reasons. First, as shown in Eq.\ref{eq:nt}, increasing the sound speed will decrease the Mach number for a given $\sigma_0$ and hence decrease $n_T$. Second, the increased sound speed at low density increases the Jeans length (see Eq. \ref{eq:jeans}). The result of both is that the \NHnH is brought slightly closer to the \citetalias{Bisbas2023} fit. The most self-consistent method to include variable $c_s$ and $\mu_p$ is an iteration process between a PDR model and the analytic relationship.

\section{Power-law index of the \NHnH relationship}\label{app:plaw}
Previous works, such as \citet{VanLoo2013} and \citet{Safranek-Shrader2017}, fitted the results from their three-dimensional simulations for a polynomial expression for \NHnH in dense gas. We calculated the slope of the analytic expressions from the derivative $d\log N_{\rm eff}/d\log n_{\rm H}$ to examine the behavior in the asymptotic limits and transition regimes using the {\sc Numpy} gradient function.

\begin{figure*}[htb!]
    \centering
    \includegraphics[width=\textwidth]{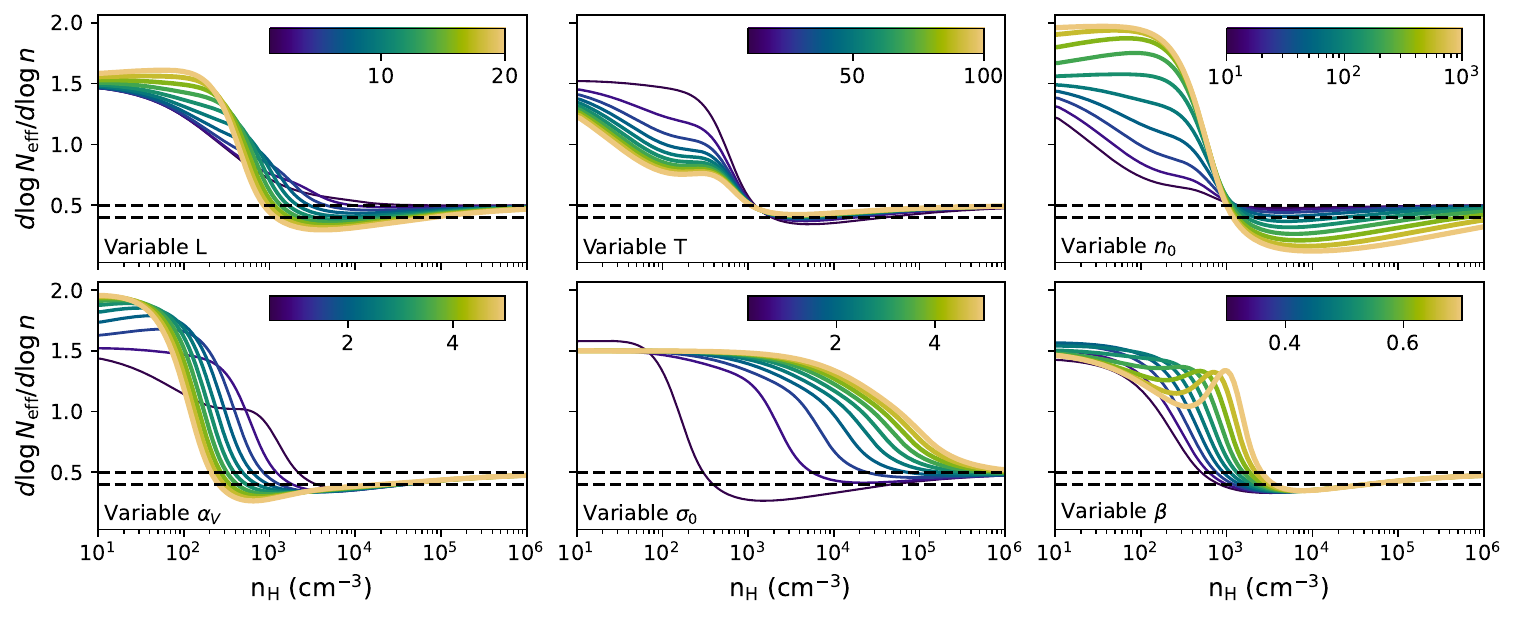}
    \caption{Same in Figure \ref{fig:NHnH_param} but showing the derivative, $d\log N_{\rm eff}/d\log n$. The horizontal lines show a slope of 0.4 and 0.5, corresponding to the results from hydrodynamical simulation fits (see text) and the isothermal Jeans-only limit.}
    \label{fig:NHnH_slope}
\end{figure*}

Figure \ref{fig:NHnH_slope} shows the computed slopes across the parameter space in Table \ref{tab:params}. All models, with the exception of high values of $n_0$, have their slopes asymptote to the isothermal Jeans-only limit of 0.5 at high density. The power-law indices at intermediate densities are highly dependent on the assumed parameters due to changes in where the transition between turbulence and gravitational regimes occurs. In models with lower $\sigma_0$, the slope reaches a minimum between $10^3 \leq n_{\rm H} \leq 10^4$, followed by a slight steepening of the slope towards 0.5. At low density, the model produces a range of superlinear power-law slopes from 1.5 to 2, with most of the solutions asymptoting around $N_{\rm eff} \propto n^{1.5}$, with $n_0$ and $\alpha_V$ being the most impactful. The average density, $n_0$, is also the most impactful at the high density asymptotic limit.

\begin{figure}[htb!]
    \centering
    \includegraphics[width=0.9\columnwidth]{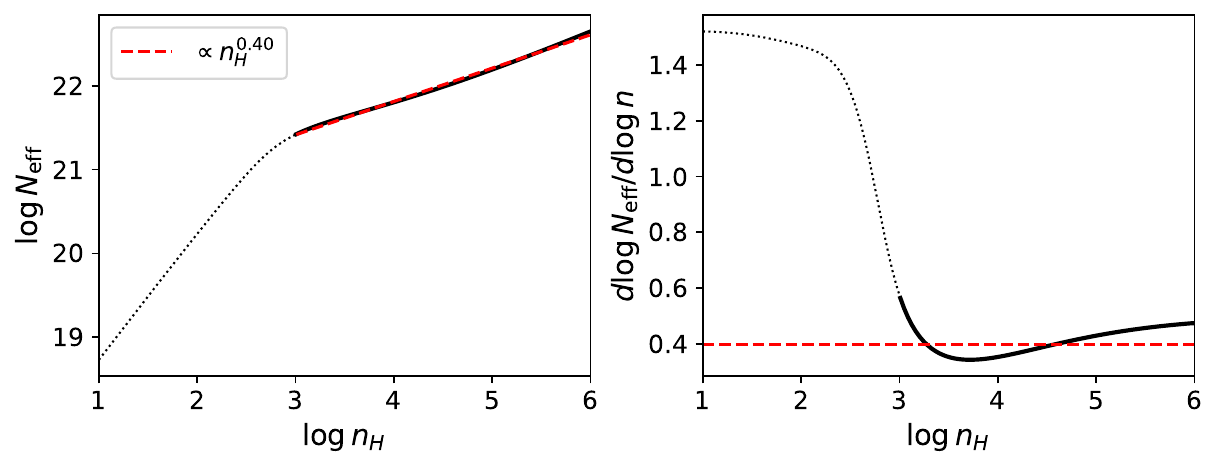}
    \caption{Comparison between the gradient calculation and a power-law fit in log-log space. Left: The \NHnH relationship for the Fiducial model (black), as shown in Figure \ref{fig:NHnH_fid}, and polynomial fit for the dense gas (red). Right: Derivative, $d\log N_{\rm eff}/d\log n$, as a function of density (black) with the power-law fit as the horizontal line. The dense gas regime with $n_{\rm H} > 10^3$ is highlighted through a bold line. The power-law relationship is annotated at the top.}
    \label{fig:gradFitCompare}
\end{figure}

The high-density power-law index asymptotes to $N_{\rm eff} \propto n_{\rm H}^{1/2}$. This is close to the slopes computed from the simulation results discussed above, which found $N_{\rm eff} \propto n_{\rm H}^{0.4}$. This is explained in Figure \ref{fig:gradFitCompare}, which compares the slope of the fiducial \NHnH to a linear fit in log-log space restricted to high density. While the constant fitted slope is consistent with previous works, the slope as a function of density varies between 0,3 and 0.6. The slightly flatter slope of the power-law fit is due to the inclusion of the turbulent column density component biasing the power-law slope fit to a flatter slope than a pure isothermal Jeans component.


\end{appendix}

\end{document}